\documentclass[twocolumn,showpacs,preprintnumbers,amsmath,amssymb,prb]{revtex4}

\usepackage{graphicx}
\usepackage{dcolumn}
\usepackage{bm}

\newcommand{\er}{{\mathbb R}}

\newcommand{\boldU}{\mathbf{U}}
\newcommand{\boldV}{\mathbf{V}}
\newcommand{\boldb}{\mathbf{b}}
\newcommand{\overU}{\overline{U}}
\newcommand{\overV}{\overline{V}}

\begin{document}


\title{Spatially Distributed Stochastic Systems: \\
equation-free and equation-assisted preconditioned computation}


\author{Liang Qiao$^1$}
\author{Radek Erban$^2$}
\author{C. T. Kelley$^3$}
\author{Ioannis G. Kevrekidis$^{1,4}$}

\affiliation{$^1$Department of Chemical Engineering, Princeton
University, Princeton, NJ 08544, USA\\
$^2$University of Oxford, Mathematical Institute, 24-29 St. Giles', Oxford, OX1 3LB, UK\\
$^3$Department of Mathematics, Box 8205, Center for Research in
Scientific Computation, North Carolina State University, Raleigh,
NC
27695, USA\\
$^4$Program in Applied and Computational Mathematics (PACM),
Princeton University, Princeton, NJ 08544, USA}

\date{\today}

\begin{abstract}
Spatially distributed  problems are often approximately modelled
in terms of partial differential equations (PDEs) for appropriate
coarse-grained quantities (e.g. concentrations).
The derivation of accurate such PDEs starting from finer scale,
atomistic models, and using suitable averaging, is often a
challenging task; {\it approximate} PDEs are typically obtained
through mathematical closure procedures (e.g. mean-field
approximations).
In this paper, we show how such {\em approximate} macroscopic PDEs
can be exploited in constructing {\em preconditioners} to
accelerate stochastic simulations for spatially distributed
particle-based process models.
We illustrate how such preconditioning can improve the convergence
of equation-free coarse-grained methods based on coarse
timesteppers.
Our model problem is a stochastic reaction-diffusion model capable
of exhibiting Turing instabilities.
\end{abstract}

\maketitle

\section{Introduction}

Many mathematical models involving reaction and diffusion (e.g.
catalytic reactions, morphogenesis) are based on partial
differential equations (PDEs) describing the evolution of species
concentrations in space and
time~\cite{Fogler:2005:CHE,Murray:2002:MB}.
One advantage of such PDE models is the extensive set of existing
theoretical and computational tools for their analysis and
efficient simulation.
A disadvantage of such continuum-based models in certain chemical
and biological contexts is the relatively low number of molecules
of some of the species involved.
This may render mean-field type PDE models inaccurate;
individual-based stochastic models become then more appropriate
than  continuum ones.

Directly using a stochastic, molecular-based model for a spatially
distributed pattern-forming system will typically be very
computationally intensive.
It becomes then important to extract useful coarse-grained,
macroscopic information from the microscopic molecular-based model
using as few detailed simulations as possible.
This is the goal of equation-free methods
\cite{Kevrekidis:2003:EFM,Gear:2002:CIB,%
Runborg:2002:EBA,Makeev:2002:CBA,Erban:2006:GRN,Salis:2005,Laing:2006,Samaey:2006}
which were designed for cases where the exact {\em macroscopic}
equations are unavailable in closed form.

Deriving accurate macroscopic equations rigorously is a
challenging task (see discussions in
Refs.~\cite{Cercignani:1988:BEA,Erban:2004:ICB,Erban:2006:TAE} for
fluid dynamics, bacteria or eukaryotic cells, respectively).
The mathematical assumptions leading to closures may not be
quantitatively correct over large parameter regimes, making the
PDE models inaccurate there.
Equation-free methods {\em circumvent} the derivation of accurate
PDEs by using short bursts of fine-scale simulation to estimate
necessary numerical quantities (residuals, action of Jacobians) on
demand, rather than through explicit closed formulas.
Here we will illustrate how even {\em approximate} PDE models can
be exploited to accelerate ``exact" (particle-based) simulations.
In the context of equation-free methods this can be accomplished
naturally by using the approximate PDEs to construct {\it
preconditioners} in the iterative numerical linear algebra
involved in fixed point, stability and bifurcation computations.
We call this procedure {\em ``equation-assisted"} computation.

Our model problem is a stochastic reaction-diffusion system
capable of exhibiting a Turing instability \cite{Turing:1952:CBM}.
Such models can serve as a prototypes of more realistic pattern
formation mechanisms during morphogenesis (see e.g.
Refs.~\cite{Shimmi:2004:FTD,Reeves:2005:CAE}).
The paper is organized as follows:
In Section \ref{secstomodel}, we introduce this illustrative
stochastic reaction-diffusion model: the Schnakenberg
\cite{Schnakenberg:1979:SCR} system of two chemical species in one
dimension.
It can predict pattern formation under some conditions and it was
also used previously in complex models of limb
development~\cite{Izaguirre:2004:CMF}.
In Section \ref{secmacroPDE} we start by presenting the mean field
Schnakenberg PDEs at the limit of large particle numbers, and
briefly summarize their bifurcation behavior in a parameter regime
where they exhibit pattern formation.
We briefly describe, in Section \ref{sectheory}, the computation
of bifurcation diagrams using a timestepper based approach -- both
deterministic and ``equation-free",  based on a stochastic
simulator implementing a spatially discretized version of the
Gillespie Stochastic Simulation Algorithm
\cite{Gillespie:1977:ESS} (SSA).
We also discuss basic features of preconditioning and
``equation-assisted" bifurcation computations.
We then present, in Section \ref{secresults}, our
``equation-assisted" results and discuss their comparison with the
``equation-free" case.
Here the mean field PDE is used to construct a preconditioner, to
accelerate the numerical linear algebra in our coarse-grained
steady state computations.
We conclude with a brief summary and discussion.

\section{The stochastic reaction-diffusion model}

\label{secstomodel}

We consider the Schnakenberg \cite{Schnakenberg:1979:SCR} system
of two chemical species $U$ and $V$ with the following reaction
mechanism:
\begin{eqnarray}
&A \quad \xrightarrow{k_1} \quad U \quad
\xrightarrow{k_2} \quad C \label{Ureact}\\
&B \quad \xrightarrow{k_3} \quad V \label{Vreact}\\
&2 U + V \quad \xrightarrow{k_4} \quad 3 U. \label{UVreact}
\end{eqnarray}
Here Eq.~(\ref{Ureact}) describes production and degradation of
$U$ and Eq.~(\ref{Vreact}) describes production of $V$.
Moreover, $U$ is also produced in the reaction
Eq.~(\ref{UVreact}).
We will assume that the concentrations of $A$ and $B$ are
constants.

To simulate stochastically Eqs.~(\ref{Ureact}) -- (\ref{UVreact}),
one can use the Gillespie SSA, a standard way to model
stochastically a spatially homogeneous (well mixed) chemical
system.
The algorithm is based on answering two essential questions at
each time step: when will the next chemical reaction occur, and
what kind of reaction will it be?
Gillespie \cite{Gillespie:1977:ESS} derived a simple way to answer
these two questions -- at each step, the computer performs a
reaction, updates numbers of reactants and products and continues
with another time step until the algorithm reaches a time of
interest.

Next, we introduce diffusion to the system.
We assume that $U$ diffuses with (macroscopic) diffusion
coefficient $\overline{d_1}$ and $V$ diffuses with (macroscopic)
diffusion coefficient $\overline{d_2}$.
We consider a spatially one-dimensional domain -- the interval
$[0,1]$ with suitable boundary conditions as specified later.
The generalization of Gillespie's ideas to spatially
nonhomogeneous systems can be found in the literature (see e.g.
Refs.~\cite{Stundzia:1996:SSC,Isaacson:2006:IDC}).
Here, we follow the most straightforward way, adding diffusion as
another set of ``reactions" to the system.
Namely, we divide our domain into $m$ boxes (small intervals) of
length $h=1/m$.
We denote by $U_i$ and $V_i$ the number of respective molecules in
the spatial interval $[(i-1)/m,i/m]$ for $i=1,\dots,m$.
This means that we describe the state of the stochastic
reaction-diffusion system by two $m$-dimensional vectors $\boldU =
\big[ U_1, U_2 \dots, U_m \big],$ $\boldV = \big[ V_1, V_2 \dots,
V_m \big]$ and we consider the following reactions at each time
step
\begin{equation}
\left.\begin{array} {l}
A~\xrightarrow{k_1}~U_i~\xrightarrow{k_2}~C,
\\B~\xrightarrow{k_3}~V_i,\\2U_i +
V_i~\xrightarrow{k_4}~3U_i,
\end{array}
\right\}~~i=1, \dots, m, \label{reactboxi}
\end{equation}
\begin{align}
U_i~&\mathop{\longrightarrow}^{d_1} U_{i+1},~~V_i
\mathop{\longrightarrow}^{d_2} V_{i+1}, ~~~~~~~~i=1, \dots, m-1,
\label{diffip1}\\
U_i~&\mathop{\longrightarrow}^{d_1} U_{i-1},~~V_i
\mathop{\longrightarrow}^{d_2} V_{i-1},~~~~~~~~i=2, \dots, m,
\label{diffin1}
\end{align}
where Eq.~(\ref{reactboxi}) means that we implement the
Schnakenberg reaction mechanism (Eqs.~(\ref{Ureact}) --
(\ref{UVreact})) in every spatial box.
Equations (\ref{diffip1}) -- (\ref{diffin1}) describe diffusion;
the transition rates between boxes are denoted by $d_1$ and $d_2$
and they are connected to the macroscopic diffusion coefficients
through the formulas $d_1 = \overline{d_1}/h^2 = \overline{d_1}
m^2$, and $d_2 = \overline{d_2}/h^2 = \overline{d_2} m^2.$
The Eqs.~(\ref{reactboxi}) -- (\ref{diffin1}), together with
suitable boundary conditions, will be simulated using Gillespie
SSA and will be our illustrative stochastic reaction-diffusion
problem in this paper.

\section{Deterministic analysis of the model problem}

\label{secmacroPDE}

\begin{figure}
\centering
\includegraphics[width=0.6\columnwidth]{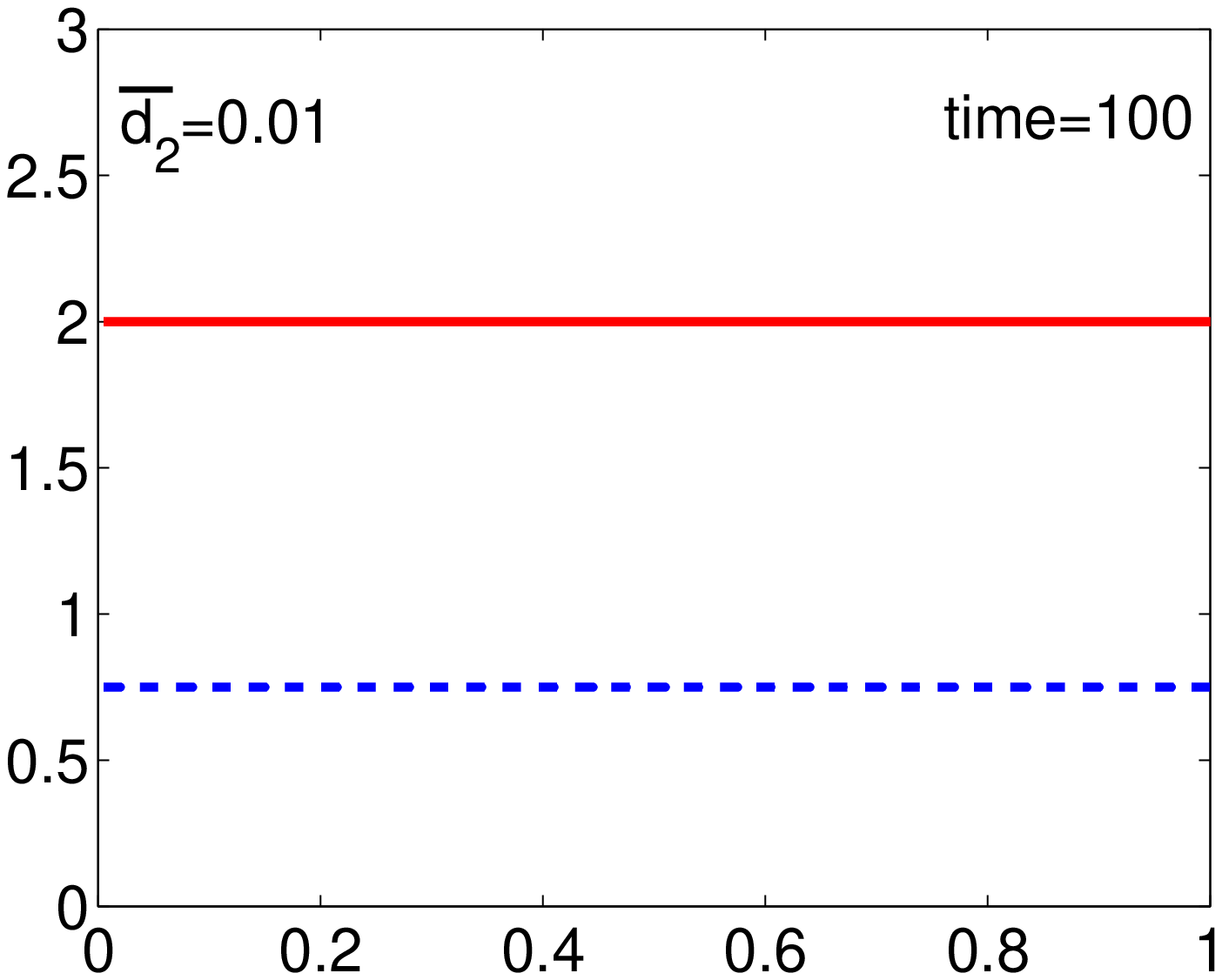}
\includegraphics[width=0.6\columnwidth]{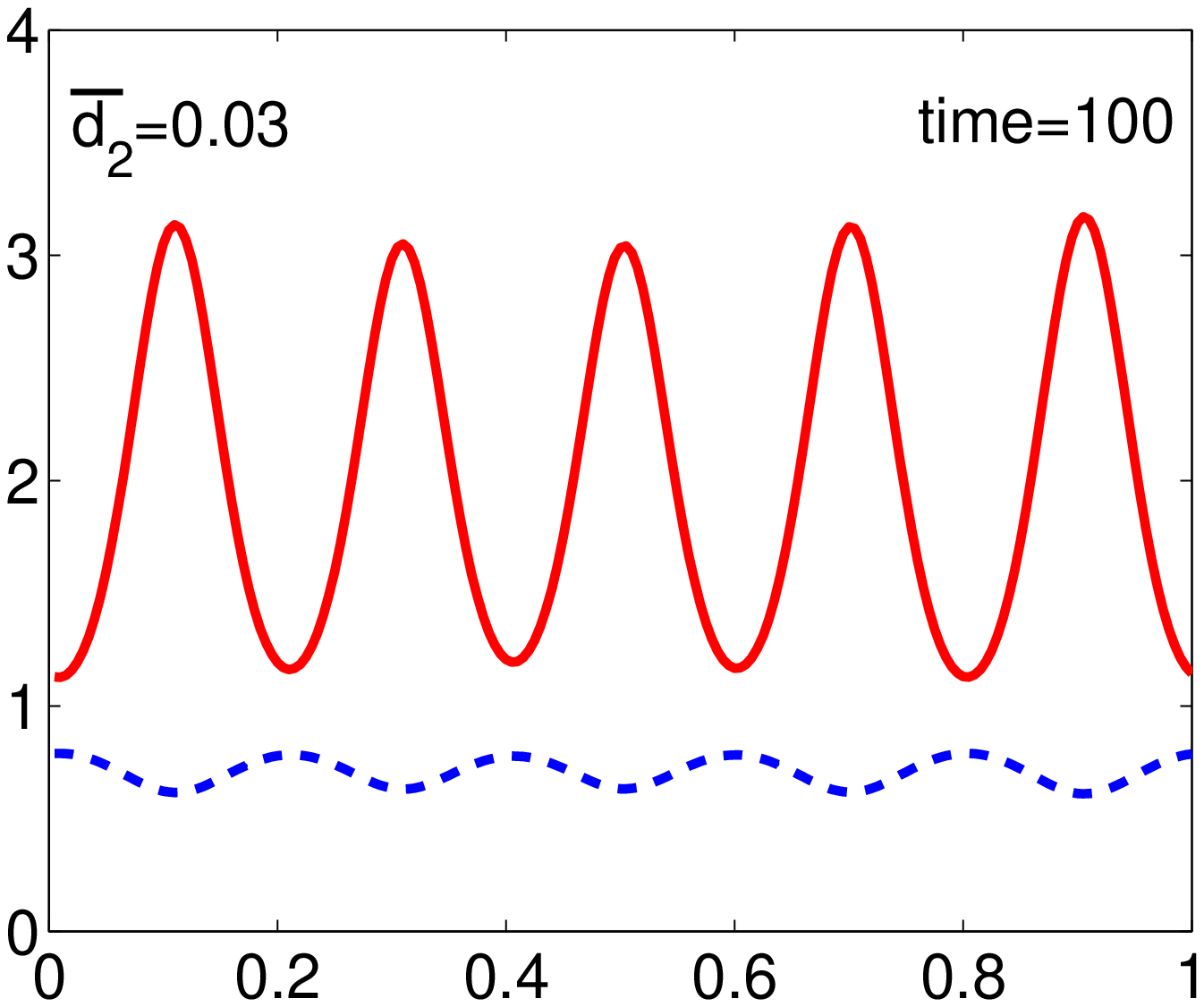}
\includegraphics[width=0.6\columnwidth]{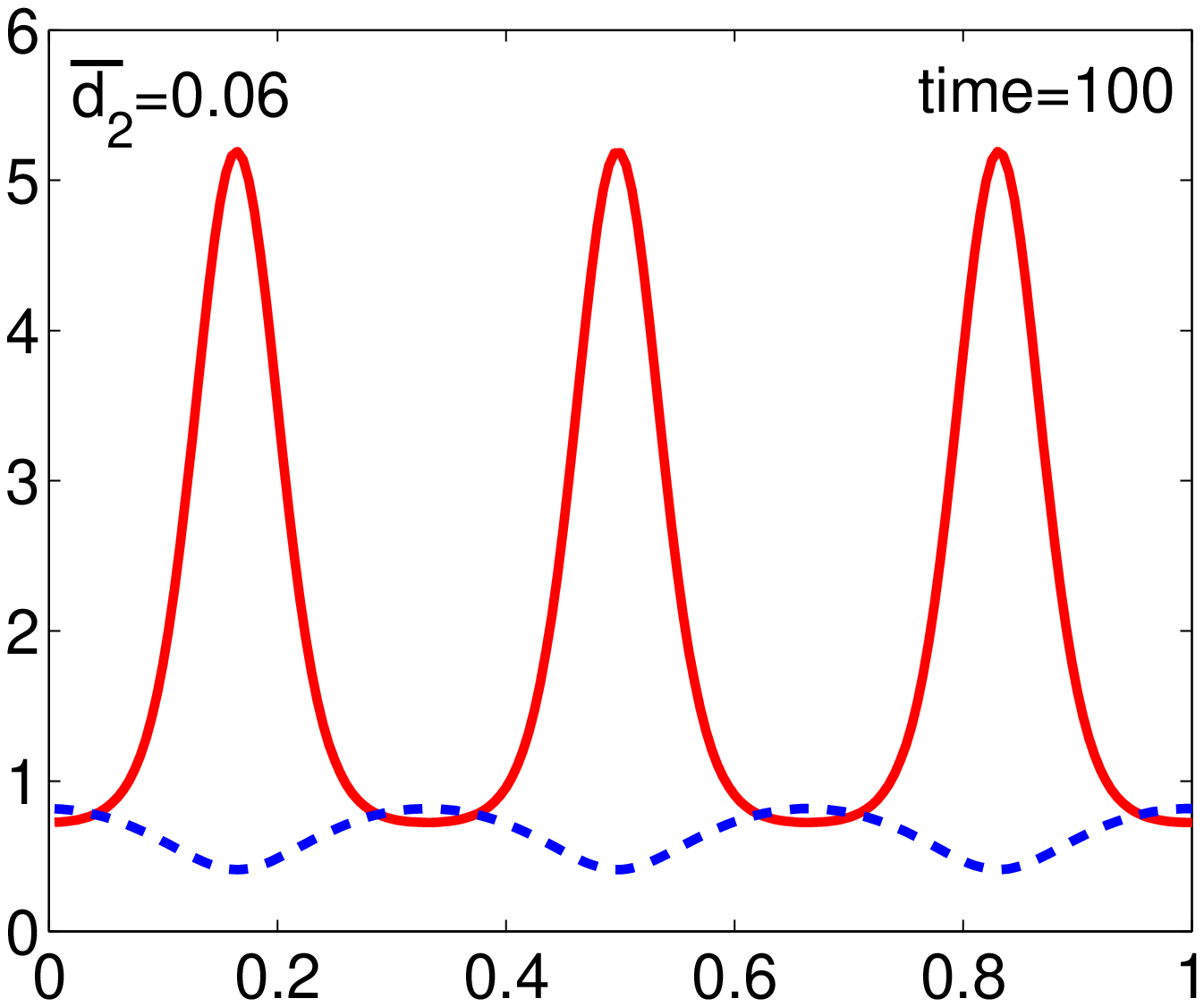}
\caption{(Color online) Solutions of Eqs.~$(\ref{rd1})$ --
$(\ref{rd2})$ with periodic boundary conditions for
$\overline{d_1} = 5 \times 10^{-4}$, $\overline{d_2} = 0.01$, 0.03
and 0.06 at $t=100$ with an initial condition being the perturbed
uniform steady state.
Spatial patterns develop for $\overline{d_2} > 0.0198$; the
solutions were shifted to have a local $\overU$ minimum at $x=0.$
We plot $\overU$ (red curve) and $\overV$ (blue curve) in the same
picture.} \label{figilldet}
\end{figure}

\begin{figure}
\centering
\includegraphics[width=0.8\columnwidth]{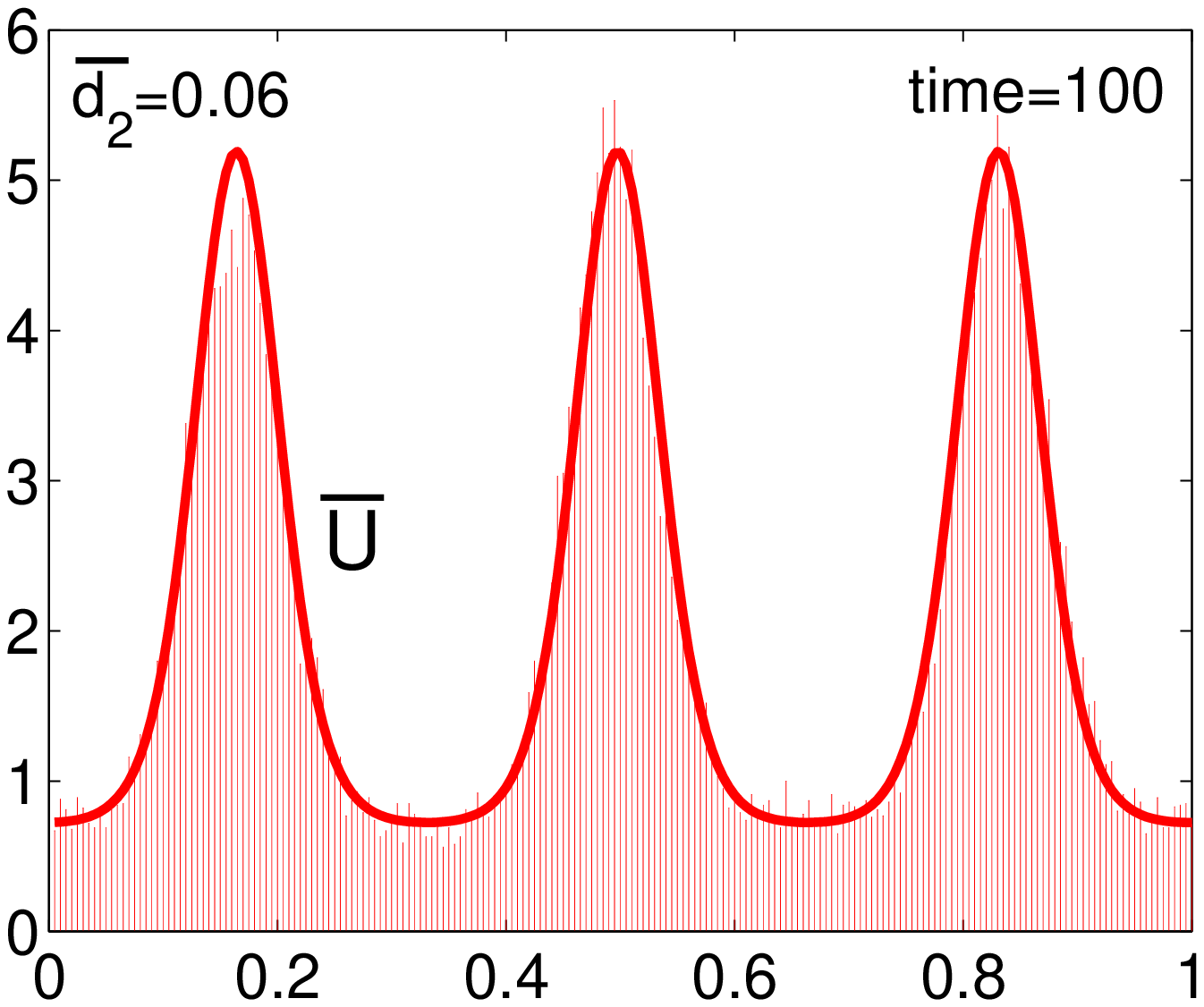}
\includegraphics[width=0.8\columnwidth]{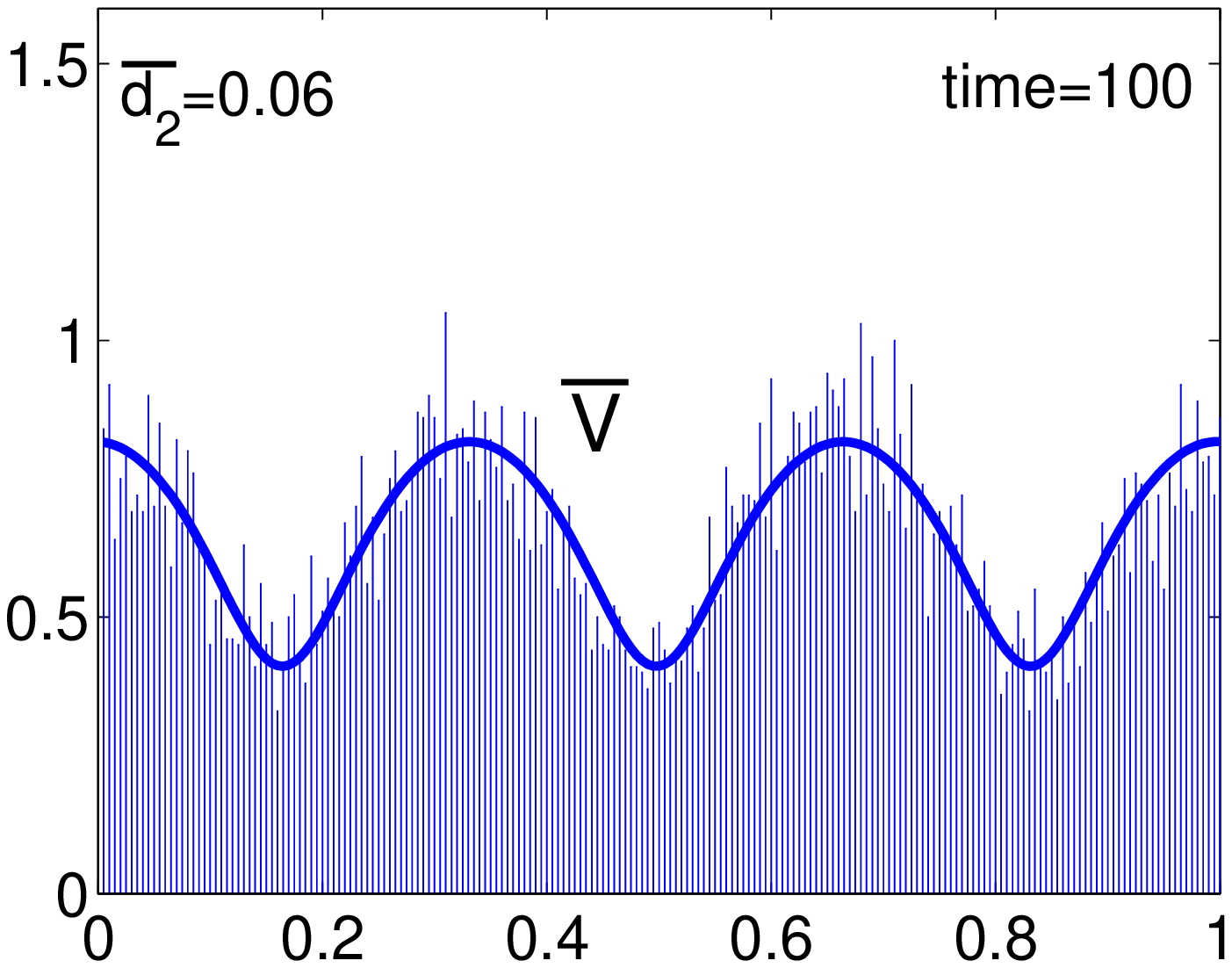}
\caption{(Color online) Comparison of histograms obtained by
stochastic simulations with the deterministic results given by
solution of Eqs.~(\ref{rd1})--(\ref{rd2}) with periodic boundary
conditions.
%
Solutions were shifted to have the local $\overU$ minimum at
$x=0.$} \label{figstodet}
\end{figure}

If we have enough molecules in the system, then
Eqs.~(\ref{reactboxi}) -- (\ref{diffin1}) are well approximated by
a system of two reaction-diffusion PDEs for the species
concentrations $\overline{U}$ and $\overline{V}$; at the mesh
points $x_i$, $ \overU(x_i) = U_i/\omega,$ $\overV(x_i) =
V_i/\omega,$ where $x_i =  \left( i - 1/2 \right) h$.
The constant $\omega$ can be interpreted as the number of
molecules in the box corresponding to a dimensionless
concentration of 1.
Reaction and diffusion rates are scaled as follows: $A = \omega
\overline{A}$, $B = \omega \overline{B},$ $k_1 = \overline{k_1},$
$k_2 = \overline{k_2},$ $k_3 = \overline{k_3},$ $k_4 =
\overline{k_4}/\omega^2,$ $d_1 = \overline{d_1}/h^2,$ $d_2
=\overline{d_2}/h^2,$ where $\overline{A}$ and $\overline{B}$ are
(constant) concentrations of reactants $A$ and $B$,
$\overline{k_j}$, $j=1,2,3,4$, are macroscopic reaction rate
constants and $\overline{d_1}$, $\overline{d_2}$ are macroscopic
diffusion coefficients.
Next (instead of further nondimensionalization) we simply choose
values for the six kinetic parameters and we study the Turing
instabilities of the resulting model.
In the rest of this paper, we put $\overline{A} = 1$, $
\overline{B} = 1$, $\overline{k_1} = 1$, $\overline{k_2} = 2$,
$\overline{k_3} = 3$, $\overline{k_4} = 1$.
Then the scaling of kinetic constants reads as follows
\begin{align}
A &= \omega, \;\; B = \omega, \;\; k_1 = 1, \;\; k_2 = 2,
\;\nonumber\\\; k_3 &= 3, \;\; k_4 = \frac{1}{\omega^2}, \;\; d_1
= \frac{\overline{d_1}}{h^2}, \;\; d_2 =
\frac{\overline{d_2}}{h^2}. \label{consca2}
\end{align}
Passing the number $\omega$ of molecules in a box to infinity and
the box length $h$ to zero, i.e. $\omega \to \infty$ and $h \to
0_+$, one can derive the following system of macroscopic partial
differential equations for concentrations $\overU$ and $\overV$.
\begin{equation}
\frac{\partial \overU}{\partial t} = \overline{d_1} \,
\frac{\partial^2 \overU}{\partial x^2} + 1 - 2 \overU + \overU^2
\, \overV \label{rd1}
\end{equation}
\begin{equation}
\frac{\partial \overV}{\partial t} = \overline{d_2} \,
\frac{\partial^2 \overV}{\partial x^2} + 3 -  \overU^2 \, \overV
\label{rd2}
\end{equation}
Here, $\overU : [0,1] \to [0,\infty)$, $\overV : [0,1] \to
[0,\infty)$ and suitable boundary conditions (e.g. no-flux,
periodic) must be introduced.
Considering no-flux or periodic boundary conditions, one can
easily verify that the homogeneous steady state of
Eqs.~(\ref{rd1}) -- (\ref{rd2}) is given by $U_h (x,t) \equiv 2,$
$V_h (x,t) \equiv 3/4$.
Linearizing Eqs.~(\ref{rd1}) -- (\ref{rd2}), one sees that the
homogeneous steady state is stable for $\overline{d_1} =
\overline{d_2} = 0$, i.e. when no diffusion is present in the
system.
In fact, the same result holds whenever $\overline{d_1} =
\overline{d_2}$: no spatial patterning can be expected if the
diffusion coefficients of both species are the same.
However, Turing \cite{Turing:1952:CBM} showed that the homogeneous
steady state $(U_h,V_h)$ might become unstable for $\overline{d_1}
\ne \overline{d_2}$.
Indeed, linearizing Eqs.~(\ref{rd1}) -- (\ref{rd2}), one finds
that the steady state $(U_h,V_h)$ is unstable and spatial patterns
develop if $\overline{d_2} > 39.6 \, \overline{d_1}$.
In this paper, we fix the diffusion coefficient $\overline{d_1} =
5 \times 10^{-4}$.
Spatial patterns may then develop for $\overline{d_2} > 0.0198$.
We show numerically computed solutions of Eqs.~(\ref{rd1}) --
(\ref{rd2}) with periodic boundary conditions for different values
of the diffusion coefficient $\overline{d_2}$ in
Fig.~\ref{figilldet}.
The initial condition was chosen to be $(U_h,V_h)$ with small
additive random noise.
The graphs of $\overU$ (red solid curve) and $\overV$ (blue dashed
curve) are plotted at dimensionaless time $t=100$, and can be
practically considered as steady states; this has been confirmed
also through a steady state solver.
%
%

%
In Fig.~\ref{figstodet}, we compare representative SSA results
with the deterministic ones.
We divide the domain $[0,1]$ into $m=200$ boxes and using
$\omega=100$ (i.e., defining density scaling so that dimensionless
density 1 corresponds to $100$ molecules of the relevant chemical
species in a box) and we choose the values of the parameters as in
Eq.~(\ref{consca2}) together with $\overline{d_1} = 5 \times
10^{-4}$ and $\overline{d_1} = 0.06$.
We will quantify the fluctuations of the stochastic simulations at
stationarity below.

\section{Equation-assisted computation: theoretical framework}\label{sectheory}

\subsection{Numerical bifurcation computations}

\begin{figure}
\centering
\includegraphics[width=0.9\columnwidth]{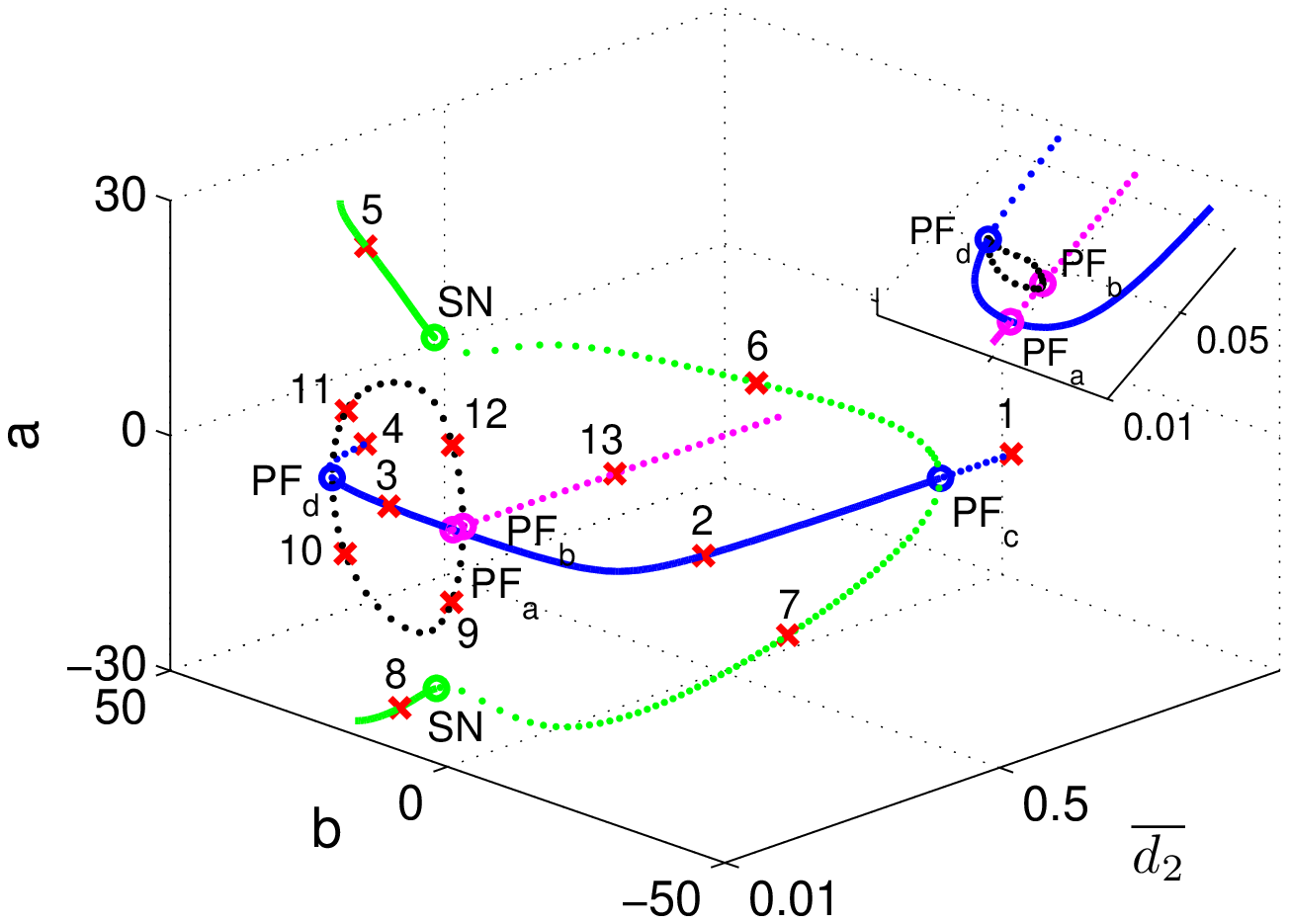}
\includegraphics[width=0.9\columnwidth]{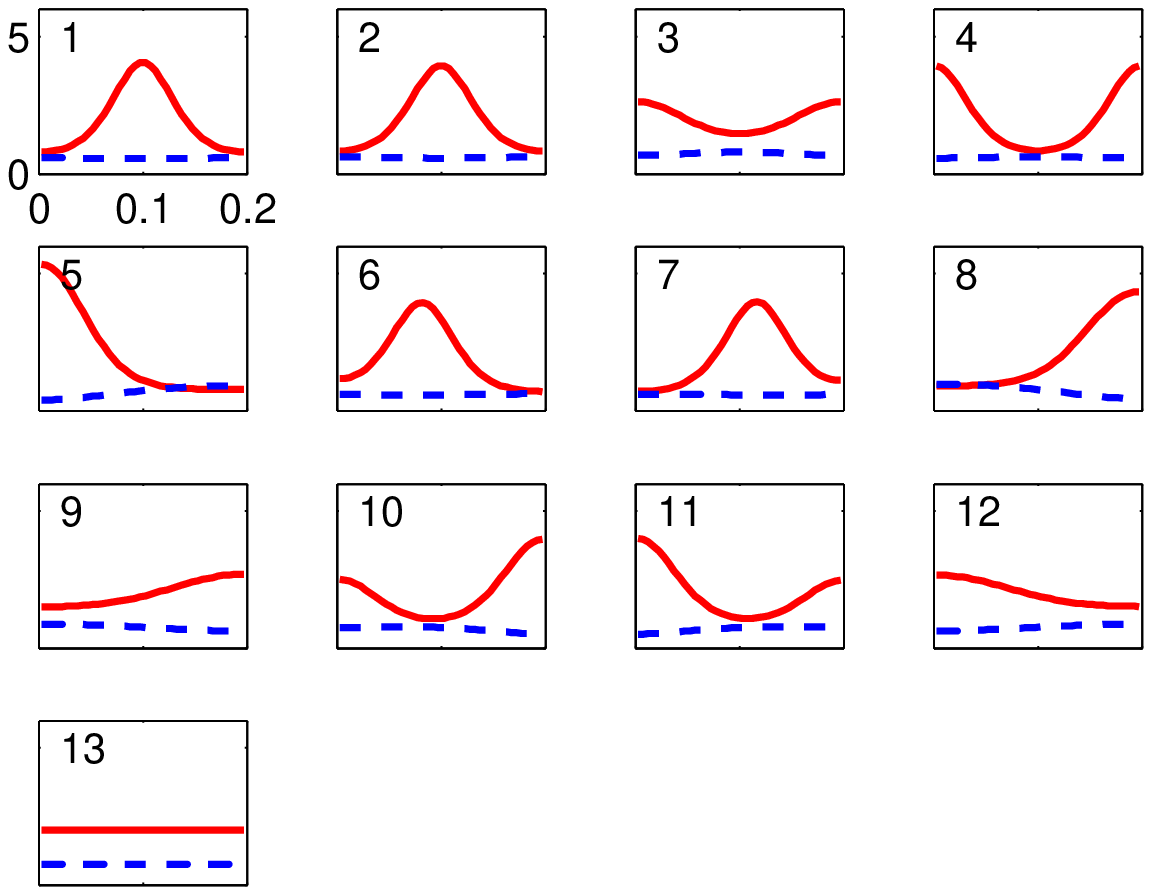}
\caption{(Color online) Bifurcation diagram of the deterministic
Schnakenberg system of two species ($U$ and $V$) with respect to
the diffusion coefficient $\overline{d_2}$.
The diffusion coefficient for $U$, $\overline{d_1}$ is fixed at $5
\times 10^{-4}$.
Axis notation: $a$ and $b$ are the first Fourier coefficients of
the solution (for $\sin(x)$ and $\cos(x)$ respectively).
Different steady state solution branches are plotted in different
colors.
Solid (dashed) lines represent stable (unstable) steady states.
The bifurcation points along the solution curves are marked by
circles and denoted with the corresponding bifurcation type (``SN"
for Saddle-Node bifurcation and ``PF" for Pitchfork bifurcation).
The straight line represents the uniform steady state (as shown in
the solution profile for point 13).
The inset is a blowup of the bifurcation diagram for small
$\overline{d_2}$.
Representative solution profiles (solid (dashed) line for $\overU$
($\overV$)) on different branches (numbered and marked by
``$\times$'') are also included.} \label{bifurdiagram}
\end{figure}

The computer-assisted study of Turing patterns in a deterministic
PDE context requires the numerical computation and parametric
continuation of steady states.
Spatially distributed PDE steady states in a bifurcation diagram
are, in general, computed by discretizing the PDE into a
(sufficiently) large set of ODEs of the type $d\mathbf
x/dt=\mathbf f(\mathbf x;\mathbf p)$, finding the roots of
$\mathbf f(\mathbf x;\mathbf p)=\mathbf 0$ and continuing them in
parameter space.
Here $\mathbf x\in \er^{N}$ is a vector containing the system
state (the discretized concentrations of $U$ and $V$), $\mathbf f:
\er^{N}\to\er^{N}$ is the right hand side of the discretization of
Eqs.~(\ref{rd1}) and~(\ref{rd2}) and $\mathbf p\in \er^M$ is a
$M$-dimensional parameter vector; here $M=1$ since we consider the
single parameter $\overline{d_2}$.
Pseudo-arclength continuation and branch switching are by now
standard numerical tools that have been incorporated in special
purpose packages like AUTO~\cite{auto81} or
CONTENT~\cite{content}.

For deterministic problems for which a good {\it dynamic
simulator} is available, the so-called ``timestepper-based"
approach allows the computation of bifurcation diagrams in the
form of a ``wrapper" around the dynamic simulator (see e.g.
Refs.~\cite{Shroff:1993:RPM, Tuckerman:1999:TS}).
Given the current state $\mathbf x$ as an initial condition, the
timestepper computes the future (after a ``reporting time" $T$)
state $ \mathbf \Phi_T(\mathbf x;\mathbf p)\equiv \mathbf x(T) $.
Steady states of the original system are then found as fixed
points of $\mathbf \Psi(\mathbf x)=0$ where $\mathbf \Psi(\mathbf
x) \equiv \mathbf x-\mathbf \Phi_T(\mathbf x;\mathbf p)$.

The bifurcation diagram in Fig.~\ref{bifurdiagram} has been
computed in both ways (giving, of course, identical results); a
discussion of some pertinent details can be found in the Appendix.
Fixed point algorithms, like the Newton-Raphson iteration,
constitute the workhorse of these solutions of large sets of
coupled, nonlinear algebraic equations; these involve the
repeated solutions of large sets of linear, coupled algebraic
equations.
Consider now performing these repeated linear solves through {\em
matrix-free} iterative linear solvers (such as
GMRES~\cite{Kelley:1995:GMRES}); in a nonlinear equation context
we will typically use a matrix-free Newton-GMRES
solver~\cite{Kelley:1995:GMRES}.
For the timestepper-based computation (solving the nonlinear
system $\mathbf \Psi(\mathbf x)=0$) we do not need to compute the
Jacobian $\bm{\mathcal{D}}\mathbf{\Psi}\equiv \partial\mathbf
\Psi(\mathbf x)/\partial\mathbf x$.
We only need to compute matrix-vector products of this Jacobian
with given known vectors $\mathbf v$, which can be estimated by  a
finite different approximation $\bm{\mathcal{D}}\mathbf{\Psi}\cdot
\mathbf v\approx[\mathbf \Psi(\mathbf x + \varepsilon \mathbf
v)-\mathbf \Psi(\mathbf x)] /\varepsilon$ with suitably small
$\varepsilon$.

Such {\it matrix-free} linear algebra methods constitute an
important component of {\em equation-free} bifurcation
calculations.
In this context macrosocopic, coarse-grained equations are not
explicitly available; yet we believe they exist, and we do have
available a fine-scale (in this case, stochastic) dynamic
simulator.
We can then {\it substitute} the (unavailable) deterministic
timestepper with a fine scale, {\it stochastic} timestepper
involving {\it lifting}, {\it evolving}, and {\it restriction}
steps~\cite{Theod:2000:PNAS,Makeev:2002:CBA,Kevrekidis:2003:EFM}.

This provides an estimate of the (unavailable) deterministic
timestepper for the (unavailable) closed macroscopic evolution
equations, obtained {\em on demand} through the stochastic
simulator.
All computations of the matrix-free Newton-GMRES involve calls to
such a timestepper (with systematically chosen initial
conditions); the stochastic simulator can then be used to
numerically compute coarse-grained bifurcation diagrams such as
the one in Fig.~\ref{bifurdiagram} even in the absence of closed
macroscopic evolution equations (i.e., equation-free).

In this paper, we propose an ``equation-assisted" approach that
accelerates equation-free computations by linking approximate
deterministic models with accurate stochastic ones: equation-free
bifurcation computations (based on the coarse timestepper) are
{\em preconditioned} using the timestepper of an (approximate)
deterministic model.

\subsection{Preconditioning and Equation-Assisted Computation}
\label{sectpre}

Good discussions of the basic features of Newton-GMRES, as well as
pseudocode and MATLAB implementations can be found in
Refs.~\cite{Kelley:1995:GMRES,Kelley:2003}.
Consider solving a general set of $N$ nonlinear equations with $N$
unknowns, $\mathbf \Psi(\mathbf x)=0$; the linear equations to be
solved at each Newton step are of the form
\begin{equation}
{\mathbf A} {\Delta\mathbf x} = {\mathbf b} \label{precon}
\end{equation}
for ${\mathbf A} \in \er^{N \times N}$, $\boldb \in \er^N$,
${\mathbf x} \in \er^N$ with
$$
\mathbf A = \bm{\mathcal{D}}\mathbf{\Psi}\vert_{\mathbf{x=x}_c},
\quad \mathbf b =-\mathbf \Psi(\mathbf x_c),
$$
where $\mathbf x_c$ is the current solution guess at each Newton
step.
For every iterative linear solve, it is important to note that, at
each iteration in GMRES, only one call to $\mathbf \Psi(\mathbf
x)$ is needed.
GMRES {\em does not require} the Jacobian matrix
$\bm{\mathcal{D}}\mathbf{\Psi}\vert_{\mathbf {x=x}_c}$ to be
computed explicitly.
The Jacobian matrix always occurs in the form of a matrix-vector
product, which can be approximated by finite differences: $\mathbf
A\mathbf v_k\approx[\mathbf \Psi(\mathbf x_c+h\mathbf v_k)-\mathbf
\Psi(\mathbf x_c)]/h$ where $h$ is suitably small.

When Newton-GMRES is used in steady state computations using the
{\it coarse} timestepper (i.e. $\mathbf \Psi^{st}(\mathbf x)
\equiv \mathbf x-\mathbf \Phi^{st}_T(\mathbf x;\mathbf p)=0$,
where $\mathbf \Phi^{st}_T(\mathbf x;\mathbf p)$ is the coarse
timestepper based on the stochastic simulator), each evaluation of
$\mathbf \Psi^{st}(\mathbf x)$ involves evolving the coarse
timestepper $\mathbf \Phi^{st}_T(\mathbf x;\mathbf p)$ for time
$T$, which is often computationally intensive; possibly several
replica simulations need to be performed for variance reduction.
It thus becomes an important task to reduce the total number of
function evaluations to convergence.
As discussed in Ref.~\cite{Kelley:1995:GMRES}, GMRES requires less
overall function evaluations when the eigenvalues of the matrix
(i.e., $\mathbf A$ in Eq.~(\ref{precon})) are more clustered.
For a given linear system in the form of Eq.~(\ref{precon}), the
preconditioning of GMRES involves finding a regular matrix
${\mathbf P}$, such that the preconditioned linear system,
\begin{equation} {\mathbf P} {\mathbf A} {\Delta\mathbf x}
= {\mathbf P} {\mathbf b} \label{precon1}
\end{equation}
leads to a more clustered eigenvalue spectrum.
Solving Eq.~(\ref{precon}) is equivalent to solving
Eq.~(\ref{precon1}).
It is well known that the system in Eq.~(\ref{precon1}) will have
better properties (from a numerical point of view) than the
original system in Eq.~(\ref{precon}) if ${\mathbf P}$ is close to
the inverse of ${\mathbf A}$, i.e. if $\Vert {\mathbf P} -
{\mathbf A}^{-1} \Vert$ is small using a suitable matrix norm.
Hence, preconditioning by an appropriate matrix ${\mathbf P}$ can
improve the efficacy of numerical solvers for Eq.~(\ref{precon});
the goal of this paper is to show how this preconditioning idea
can be applied to equation-free stochastic reaction-diffusion
problems (and spatially distributed evolution problems more
generally).

An {\it approximate} deterministic evolution equation for the
stochastic system statistics may be available (Eqs.~(\ref{rd1})
and~(\ref{rd2})), based on closure assumptions, which is not
accurate enough to compute with; yet we can take advantage of such
an evolution equation by using it to create a ``good"
preconditioning matrix $\mathbf P$.
This preconditioning is implemented here by multiplying the
original output of each (stochastic simulation based) evaluation
of $\mathbf \Psi^{st}(\mathbf x)$ with ${\mathbf P} =
\left(\bm{\mathcal{D}}\mathbf{\Psi}^{det}\vert_{\mathbf
{x=x}_c}\right)^{-1}$, where $\mathbf\Psi^{det}(\mathbf x) \equiv
\mathbf x-\mathbf \Phi^{det}_T(\mathbf x;\mathbf p)$ is defined
using the {\it deterministic} mean field PDE timestepper $\mathbf
\Phi^{det}_T(\mathbf x;\mathbf p)$.
That is, we use the deterministic timestepper of the approximate
PDEs (here, at the current solution guess $\mathbf{x}_c$) to help
accelerate the equation-free Netwon-GMRES computation; a much
simpler preconditioning scheme would constantly use the inverse of
the deterministic timestepper {\it at the deterministic steady
state}.
At each Newton step the linear equation set to be actually solved
by GMRES after preconditioning is
$$
{\mathbf P} \, \bm{\mathcal{D}}\mathbf{\Psi}^{st}\vert_{\mathbf
{x=x}_c}\Delta \mathbf x= - \mathbf{P}
\,\mathbf{\Psi}^{st}(\mathbf x_c).
$$
For the one-dimensional problem used for illustration here, it is
easy to compute $\mathbf {Pv}$ for a given vector $\mathbf v$ by
solving ${\mathbf P}^{-1}\mathbf y=\mathbf v$ through direct
linear algebra (e.g. Gauss elimination).
In general, however, it is worth noting that  ${\mathbf
P}^{-1}\mathbf y=\mathbf v$ can be solved for $\mathbf y$ with
GMRES, through repeated calls to the (deterministic) timestepper
of the (not-so-accurate) deterministic PDEs (Eqs.~(\ref{rd1})
and~(\ref{rd2})).

\section{Results and discussion}

\label{secresults}

\begin{figure}
\centering
\includegraphics[width=0.9\columnwidth]{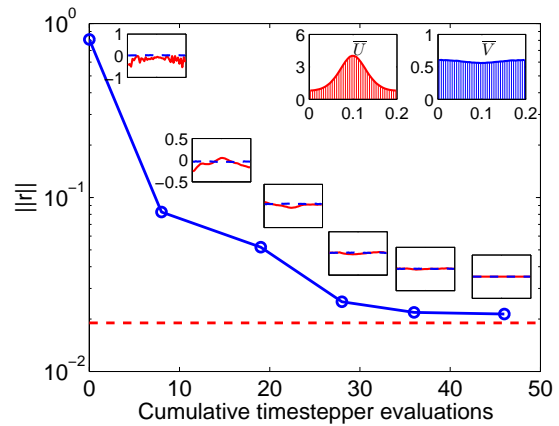}
\caption{(Color online) Convergence of Newton-GMRES based on the
coarse timestepper for a stable steady state (point 2 in
Fig.~\ref{bifurdiagram}).
The y-axis is the norm of the (coarse) residual.
The dashed line shows the estimated magnitude of the stochastic
simulator fluctuations rescaled to account for replica averaging
(which we use to estimate the fluctuations in the evaluation of
the coarse timestepper).
The relative error of the corresponding solution guess at each
Newton step is shown in the insets (solid (dashed) line for
$\overline U$($\overline V$) with y-axis ranging from -0.5 to 0.5
except the first one) along the convergence curve.
The converged coarse steady solution profiles are shown at the
upper right.
Parameters used: $m=40$ boxes in the domain $[0,0.2]$, $T=0.05$,
$\omega=2000$ with 150 copies. } \label{stable_convergence}
\end{figure}

\begin{figure}
\centering
\includegraphics[width=0.9\columnwidth]{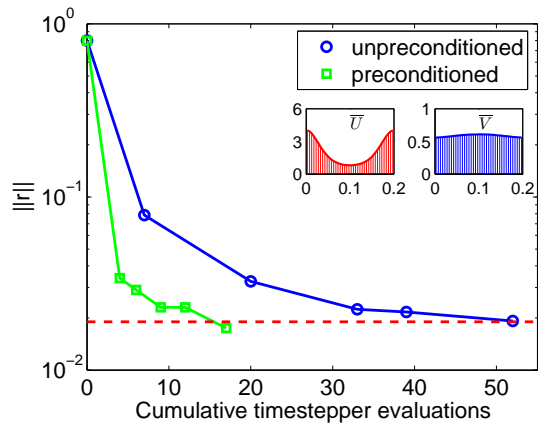}
\caption{(Color online) Convergence of Newton-GMRES based on
unpreconditioned (equation-free) and preconditioned
(equation-assisted) coarse timestepper for an {\em unstable}
coarse steady state (point 4 in Fig.~\ref{bifurdiagram}).
The dashed line shows the estimated magnitude of the stochastic
simulator fluctuations, rescaled to account for replica averaging
(which we use to estimate the fluctuations in the evaluation of
the coarse timestepper).
%
%
The upper right insets show the converged coarse steady solution
profiles.
Parameters used: $m=40$ boxes in the domain $[0,0.2]$, $T=0.05$,
$\omega=2000$ with 150 copies.} \label{preconditioning}
\end{figure}

\begin{figure}
\centering
\includegraphics[width=0.9\columnwidth]{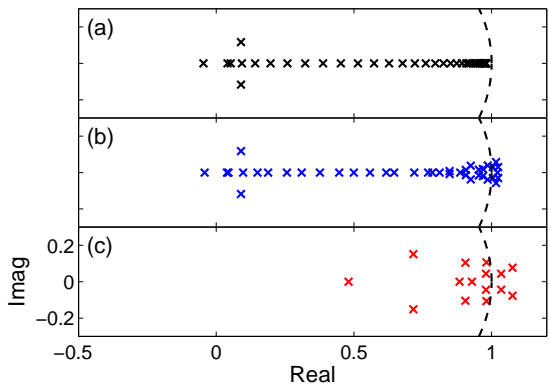}
\caption{(Color online) Comparison of the leading coarse
eigenvalues of Jacobian matrices based on deterministic
timestepper $\bm{\mathcal{D}}\mathbf{\Psi}^{det}$, the coarse
timestepper $\bm{\mathcal{D}}\mathbf{\Psi}^{st}$, and the coarse
timestepper after preconditioning
$\mathbf{P}\bm{\mathcal{D}}\mathbf{\Psi}^{st}$, which are all
evaluated at the computed unstable coarse steady state.
(a) Leading 40 (smallest magnitude) eigenvalues of
$\bm{\mathcal{D}}\mathbf{\Psi}^{det}$. The eigenvalues have
already started clustering at 1.
(b) Leading 40 eigenvalues of
$\bm{\mathcal{D}}\mathbf{\Psi}^{st}$.
This (partial) spectrum is very similar to the spectrum in (a).
The appearance of complex conjugate eigenvalue pairs close to 1
and eigenvalues larger than 1 as in (b) (and possibly also in (c))
is probably caused by the relatively large fluctuations in the
evaluation of the coarse timestepper (while the differences
between the eigenvalues within this cluster are relatively small).
(c) Leading 15 eigenvalues of $\mathbf
P\bm{\mathcal{D}}\mathbf{\Psi}^{st}$.
Most of the eigenvalues are clustered close to 1.}
\label{eig_spectrum}
\end{figure}

For our coarse timestepper (based on SSA simulation), we
discretize the one-dimensional domain [0, $L$] with $L=0.2$ into
$m=40$ equally spaced boxes.
We choose $\omega = 2000$, which means that the unit density in
each box corresponds to 2000 molecules.
Each evaluation of the coarse timestepper corresponds to evolving
150 replicas of the SSA simulator for time $T=0.05$; the average
of all replicas is reported.

We have assumed that there exist some ``underlying PDEs" that
describe the evolution of the (statistics of the) SSA simulator
averaged over an infinite number of copies.
In our computations we estimate the coarse timestepper of these
``underlying PDEs" by averaging over several (here 150) copies;
even though averaging reduces their variance, fluctuations will
always remain when a finite number of copies is used.
Note that we should not, in general, expect these ``underlying
PDEs" to be the same as the mean field PDE system described in
Section~\ref{secmacroPDE} (Eqs.~(\ref{rd1}) and~(\ref{rd2})),
which corresponds to the limit of {\it infinite numbers of
molecules}.
We do, however, know that if the parameter $\omega$ increases to
infinity, the ``underlying PDEs" do converge to Eqs.~(\ref{rd1})
and~(\ref{rd2}).

We start by using matrix-free Newton-GMRES to compute
representative spatially nonuniform (both stable and unstable)
coarse steady states through the SSA simulator.
The Newton-GMRES fixed point solver used is adapted from the
MATLAB code {\em nsoli}~\cite{Kelley:2003} with two modifications:
(a) constant relative tolerance for each GMRES solve; and (b)
since our preconditioner changes every time we update the current
solution guess, an additional function for constructing the
updated preconditioner was included as an additional input
parameter.
%
%
The convergence of Newton-GMRES to a coarse, spatially nonuniform,
stable steady state (point 2 in Fig.~\ref{bifurdiagram}) is shown
in Fig.~\ref{stable_convergence}.
The magnitude of the fluctuations introduced by the stochastic
simulator is estimated at this coarse steady state as follows:
from long-term SSA simulations, after the stationary state has
been reached, we estimate the standard deviation  $\sigma_i$ for
each $U_i$ and $V_i,$ $i= 1 \dots m$; our estimate of the
magnitude of the fluctuations is the Euclidean norm of this vector
of standard deviations.
Averaging over $n$ replicas should scale this estimate by a factor
of $\sqrt{n}$; the resulting estimate is marked by a dashed line
in Figs.~\ref{stable_convergence} and~\ref{preconditioning}.
This number also provides an estimate of a reasonable expected
residual norm upon convergence of the Newton-GMRES.

Because of the presence of fluctuations in the evaluation of the
coarse timestepper, an {\it inexact} matrix-vector product is
computed at each GMRES iteration.
The convergence of GMRES in the presence of noise is the focus of
extensive study in the current literature (see e.g.
Refs.~\cite{Valeria:2003:IGMRES,Jasper:2004:IGMRES,Amina:2005:RLX}).
These references include discussions of bounds of the attainable
accuracy of the computed solution and possible relaxation
strategies in the presence of noise.
In the context of Newton-GMRES, the right hand side of the linear
equation we want to solve at each Newton step (i.e. $-\mathbf
\Psi(\mathbf x_c)$) is also computed with fluctuations.

A representative {\em unstable} coarse steady state (point 4 in
the bifurcation diagram) is also computed with Newton-GMRES.
In this case, however, we also implemented the equation-assisted
preconditioning of the coarse GMRES, as discussed in
Section~\ref{sectheory}, using the inverse of the corresponding
Jacobian computed from the known deterministic approximate PDEs.
The convergence of Newton-GMRES before and after preconditioning
(that is equation-free and equation-assisted, respectively) is
compared in Fig.~\ref{preconditioning}.
The leading parts of the eigenvalue spectra of
$\bm{\mathcal{D}}\mathbf{\Psi}^{det}$,
$\bm{\mathcal{D}}\mathbf{\Psi}^{st}$ (the first 40 eigenvalues)
and $\mathbf P\bm{\mathcal{D}}\mathbf{\Psi}^{st}$ (the first 15
eigenvalues) evaluated at the coarse steady state are computed
using the coarse timestepper and the iterative eigenvalue solver
ARPACK (implemented in MATLAB as function {\it eigs}) and shown in
Fig.~\ref{eig_spectrum}.
A quick inspection of the numerically computed leading spectra
shows that, after preconditioning, the eigenvalues of the
preconditioned Jacobian based on the coarse timestepper were
indeed more clustered close to 1, consistent with the reduction in
GMRES iterations observed in Fig.~\ref{preconditioning}.

Note that the first vector in the Krylov subspace constructed for
GMRES  ($\mathbf v_1$) is generally obtained by setting the
initial solution guess for the linear system to zero.
This implies that $\mathbf v_1$ is {\em different} for the
preconditioned and unpreconditioned GMRES ($\mathbf b$ and
$\mathbf P \mathbf b$ for the linear system Eq.~(\ref{precon}),
respectively, where $\mathbf P$ is the preconditioner).
Since all the subsequent vectors in the Krylov subspace are built
upon the first ones, this may also lead to a difference in the
number of GMRES iterations to convergence.

Figure~\ref{preconditioning} shows the cumulative calls to the
coarse timestepper needed to reduce the {\em nonlinear residual},
which is similarly defined for both the preconditioned and the
unpreconditioned case.
The results indicate that the preconditioner is effective in
reducing the nonlinear residual, and efficient in terms of overall
timestepper evaluations.
The preconditioned and unpreconditioned {\em linear} residuals, on
the other hand, are measured in different norms ($\lVert\mathbf
P(\mathbf b-\mathbf A\Delta\mathbf x_k)\rVert_2$ and $\lVert
\mathbf b-\mathbf A\Delta\mathbf x_k\rVert_2$); we impose the same
{\em relative} tolerance for convergence.
For non-noisy problems and very tight relative termination
tolerances, the results of the two types of linear solve at the
end of the first Newton step would be practically the same; with
larger termination tolerances, given the presence of noise, this
is clearly not the case.
%
%
%
%
When the initial guess is far away from the true solution (at the
first Newton step) the initial tolerance for GMRES can be set
relatively high, to avoid ``oversolving" the linear equation at
the early stages of convergence.
%
%

\section{Conclusion}
The purpose of this paper is to illustrate a simple idea: that
coarse-grained, macroscopic equations can be used to assist
detailed, fine scale stochastic simulations even when they are not
really accurate.
This is accomplished by using certain features of such closed-form
macroscopic equations (such as their discretized linearizations)
as {\it preconditioners} in equation-free iterative linear algebra
computations.
This is then an ``equation-assisted" approach: we compute with a
coarse timestepper based on the fine scale model, but accelerate
the convergence of these computations using ``the best available"
continuum deterministic model.

In this paper we illustrated the concept using a coarse
timestepper based on a {\it spatially distributed} SSA
reaction-diffusion implementation of the Schnakenberg kinetic
scheme, and preconditioning with the corresponding Jacobian
derived from the mean-field PDEs.
This allowed us to accelerate the equation-free computation of
both stable and unstable spatially structured reaction-diffusion
steady states.
The approach can be used as a computational ``wrapper" around
different types of inner stochastic simulators.
The inner simulator was based on spatially discretized SSA; the
approach could also be wrapped around ``already accelerated" SSA
schemes (e.g. those exploiting separation of time
scales~\cite{Rao:2003:SCK,Haseltine:2002:ASC,Cao:2005:SCS}).
The approach could also be wrapped around non-SSA, lattice gas
spatially distributed kinetic Monte Carlo simulators, or around
simulators based on ``the best available" analytically
coarse-grained models of kinetic Monte Carlo processes (e.g.
Refs.~\cite{Katsoulakis:2003:PNAS,Katsoulakis:2003:JCP}).
It can also be wrapped around different (non-kMC) types of fine
scale or hybrid models such as Lattice-Boltzmann inner simulators
\cite{Samaey:2006} (with density PDE preconditioning), or around
molecular, Brownian or dissipative particle dynamics simulators of
condensed matter problems, with the preconditioning coming from
traditional continuum closures (elasticity theory, non-Newtonian
rheology).
Beyond steady state computations, such preconditioning might also
be helpful in other coarse-grained computations involving
matrix-free iterative linear algebra, such as implicit coarse
integration schemes.

\section{acknowledgement} \label{ack}
This work was partially supported by the U.S. Department of Energy
(IGK, LQ through PPPL) and DARPA, by the NSF (DMS-0404537, CTK)
and by the Biotechnology and Biological Sciences Research Council,
University of Oxford and Linacre College, Oxford (RE). It is a
pleasure for the authors to acknowledge helpful suggestions by
Giovanni Samaey and Wim Vanroose during the preparation of this
manuscript.

\appendix*
\section{The bifurcation diagram}
\label{append}

%
The bifurcation diagram with respect to the diffusion coefficient
$\overline{d_2}$ of species $V$ is computed using both the steady
state and the deterministic timestepper approach with identical
results and is plotted in Fig.~\ref{bifurdiagram}.

We used the following parameters in our computations with the
deterministic timestepper from the discretized ODE system: domain
length $L=0.2$, number of nodes $m=40$, time reporting horizon
$T=1$.
%

The stability of the steady state solutions is identified by
checking the leading eigenvalues ($\lambda_i$) of the Jacobian
matrix of the linearized ODE system evaluated at the steady
states; we confirmed that the eigenvalues of the linearization of
our timestepper at steady state ($\mu_i$) indeed satisfy
$\lambda_i = \ln\mu_i/T$.
The matrix-free eigenvalue solver ARPACK is used to compute the
leading eigenvalues for both the deterministic and the coarse
timsteppers.

The first two Fourier coefficients of the steady state solution,
$a$ and $b$ (for $\sin(x)$ and $\cos(x)$ respectively), are
plotted versus the bifurcation parameter $\overline{d_2}$.
The (partial) bifurcation diagram consists of four different
branches of solutions (plotted in different colors).
%
%
%
%
%
%

The steady states computed in this bifurcation diagram show at
most one peak due to the relatively short domain length.
Steady states with $n$ peaks (as shown in Fig. \ref{figilldet})
can be easily obtained by using $n$ copies of the one-peak
solution as building blocks; yet the stabilities of the multi-peak
and one-peak steady states are not the same (see e.g.
Ref.~\cite{Yannis:1990:SIAM}).


\end{document}